\documentstyle[12pt,aaspp4]{article}
\begin{document}

\parindent=1.0cm

\title{Deep Near-Infrared Imaging of a Field in the Outer Disk of M82 
with the ALTAIR Adaptive Optics System on Gemini North}

\author{T. J. Davidge}

\affil{Herzberg Institute of Astrophysics,
\\National Research Council of Canada, 5071 West Saanich Road,
\\Victoria, B.C. Canada V9E 2E7\\ {\it email: tim.davidge@nrc-cnrc.gc.ca}}

\author{J. Stoesz}

\affil{Department of Physics \& Astronomy, University of Victoria, 
\\Victoria, B.C. Canada V8W 3P6\\ {\it email: jeff.stoesz@nrc-cnrc.gc.ca}}

\author{F. Rigaut}

\affil{Gemini Observatory, 670 North A'ohoku Place, 
\\Hilo, HI 96720-2700\\ {\it email: frigaut@gemini.edu}}

\author{J.-P. Veran}

\affil{Herzberg Institute of Astrophysics
{\it email: jean-pierre.veran@nrc-cnrc.gc.ca}}

\author{G. Herriot}

\affil{Herzberg Institute of Astrophysics
{\it email: glen.herriot@nrc-cnrc.gc.ca}}

\begin{abstract}

	Deep $H$ and $K'$ images, recorded with the ALTAIR adaptive optics 
system and NIRI imager on Gemini North, are used to probe the red stellar 
content in a field with a projected distance of 1 kpc above the disk plane 
of the starburst galaxy M82. The data have an angular resolution of 0.08 arcsec 
FWHM, and individual AGB and RGB stars are resolved. The AGB extends to at 
least 1.7 mag in $K$ above the RGB-tip, which occurs at $K = 21.7$. 
The relative numbers of bright AGB stars and RGB stars are consistent with 
stellar evolution models, and one of the brightest AGB stars has an 
$H-K$ color and K brightness that is consistent with it being a C star. 
The brightnesses of the AGB stars suggest that they formed during intermediate 
epochs, possibily after the last major interaction with M81. Therefore, star 
formation in M82 during intermediate epochs may not have been restricted to 
the plane of the disk.

\end{abstract}


\section{INTRODUCTION}

	The M81 group is one of the nearest collections of galaxies outside of 
the Local Group. Gas bridges link many of its members 
(e.g. van der Hulst 1979; Yun, Ho, \& Lo 1994; Boyce et al. 2001), and 
simulations indicate that there were large-scale tidal interactions within the 
past few $10^8$ yr (Brouillet et al. 1991; Yun et al. 1994). These interactions 
triggered large-scale star formation in some galaxies, with 
star-formation rates exceeding those in normal disks (e.g. 
Walter et al. 2002). The M81 system thus appears to be experiencing a 
phase of rapid evolution. 

	M82 is an amorphous, nearly edge-on galaxy 
that is the third brightest member of the M81 group. M82 was 
likely an Sc/Irr galaxy at some point in the past (O'Connell \& Mangano 1978). 
However, gas streams now link M82, M81, and NGC 3077 (e.g. Yun et al. 
1994; Boyce et al. 2001), indicating that these galaxies interacted, 
spurring bursts of star formation that profoundly effected all 
three galaxies. The interactions appear to have 
triggered the formation of a population of compact, massive star clusters in 
the disk of M81 that are more typically seen in later-type 
spirals (Chandar, Tsvetanov, \& Ford 2001), while much of the 
neutral hydrogen associated with NGC 3077 is in a tidal arm that is well 
offset from the main body of the galaxy (e.g. Walter et al. 
2002; Yun et al. 1994). M82 continues to experience 
the after-effects of these interactions, as the star 
formation rate in this galaxy is currently an order of magnitude larger 
than in M81, and is almost two orders of magnitude larger than in NGC 3077 
(Ott, Martin, \& Walter 2003). The star-forming activity in M82 
drives a bipolar wind that feeds gas and dust into the outer regions of the 
galaxy (e.g. Bland \& Tully 1988), and the 
time required for this wind to deplete the disk ISM of M82 
is 140 Myr (Martin 1998). Line emission, powered by photoionization near the 
disk, and shocks at larger radii, has been traced out to 2 kpc into the 
halo of M82 (Shopbell \& Bland-Hawthorn 1998). 

	Most stellar content studies have focused on the disk of M82, and 
evidence for a large population of hot stars (e.g. O'Connell \& Mangano; 
Telesco \& Harper 1980) and supernova remnants (e.g. Kronberg, Biermann, \& 
Schwab 1981) was found early-on. The young stellar content of M82 is 
concentrated in compact star clusters, with the main body of the disk being 
largely free of recent star formation (O'Connell \& Mangano 1978). Studies of 
Region B, which was an area of concentrated star formation in the past, 
with the HST have found a large number of clusters, the age 
distribution of which peaks near log(t$_{yr}$) = 9.0 (de Grijs, 
O'Connell, \& Gallagher 2001; de Grijs, Bastian, \& Lamers 2003; Parmentier, 
de Grijs, \& Gilmore 2003)

	There are hints that recent star formation in M82 may not have been 
restricted to the disk plane. Sakai \& Madore (1999) used $V$ and $I$ WFPC2 
images to investigate the stellar content in the outer disk and inner halo of 
M82. They found a large red giant branch (RGB) population, as well as a 
number of sources brighter than the RGB-tip. While it is tempting to associate 
the brighter sources with AGB stars, Sakai \& Madore (1999) caution 
that they could be blends of fainter stars. 

	The possible detection of bright AGB stars by Sakai \& Madore (1999) is 
of interest, as the outer regions of galaxies are not expected to be sites of 
active star formation because the collapse instabilities that trigger such 
activity are expected to occur most readily in high density environments 
(e.g. Kennicutt 1989). Nevertheless, stochastic events, such as spiral density 
waves, could trigger star formation in regions where the density is otherwise 
not high enough (e.g. Ferguson et al. 1998), and intermediate-age stars are 
seen in the outer regions of the Sc galaxies NGC 2403 and M33 (Davidge 2003a), 
and the outer disk of M31 (Ferguson \& Johnson 2001). Interactions 
between the bipolar wind and molecular clouds in the 
halo of M82 may also trigger star formation at large 
distances above the disk of this galaxy. It is also possible that any young and 
intermediate-age stars in the outer regions of M82 may not have formed {\it in 
situ}, as interactions with M81 and NGC 3077 may have created tidal arms that 
moved stars out of the plane of the disk. Finally, M82 has a 
central bar (Telesco et al. 1991) that contains $20 - 40\%$ 
of the total galaxy mass (Greve et al. 2002), and the disruption of this 
structure could scatter young stars away from the center of the galaxy (e.g. 
Mayer et al. 2001). 

	High angular-resolution observations in the near-infrared can be used 
to determine if an intermediate age population is present 
outside of the central regions of M82. The brightest AGB stars, 
which probe the star-forming history of systems during 
intermediate epochs, have cool temperatures and so the 
near-infrared is a prime wavelength region for detecting these objects. In 
fact, the AGB forms a near-vertical sequence on infrared color-magnitude 
diagrams (CMDs), making it easier to detect AGB-tip stars than in visible 
wavelength CMDs, where line blanketing can cause the AGB to bend over, 
thereby placing the AGB-tip at a brightness where crowding may be significant. 
In addition, broad-band infrared photometry can be used 
to distinguish between oxygen-rich K and M giants and C stars (e.g. Davidge 
2003b), which are classical indicators of intermediate age populations.

	In the present study, deep $H$ and $K'$ images obtained with the ALTAIR 
adaptive optics (AO) system and NIRI imager on the Gemini North (GN) telescope 
are used to investigate the bright red stellar content of a field near the edge 
of the M82 disk. Field selection was based on the availability of a 
bright ($R < 12$) natural guide star reasonably close to M82. 
The guide star selected here allows a region that 
is 1 arcmin off of the plane of the disk, which corresponds to a 
linear distance of roughly 1 kpc, to be probed. Details of the 
observations and their reduction are given in \S 2, while the CMDs and 
luminosity functions (LFs) constructed from these data are discussed in 
\S 3. A brief discussion and summary of the results follows in \S 4.

\section{OBSERVATIONS AND REDUCTIONS}

	The data were obtained on the nights of May 12 and 15 2003 as part of 
the commissioning program for the ALTAIR AO system. ALTAIR is a natural guide 
star AO system with the core components being a 177 element deformable mirror 
and a $12 \times 12$ element Shack-Hartmann CCD wavefront sensor. The system 
was designed and developed at the Herzberg Institute of Astrophysics for use on 
GN. The deformable mirror is conjugated to an altitude of 6.5 km above the 
summit, which C$_{n}^2$ measurements indicate is the effective height of 
turbulence at this site (Racine \& Ellerbroeck 1995). Herriot et al. (2000) 
give a more complete discussion of ALTAIR and its design. The data were 
recorded with NIRI, which is the facility IR imager on GN. NIRI was used in 
f/32 mode for these observations, and the image scale is 0.022 arcsec 
pixel$^{-1}$; the $1024 \times 1024$ InSb array in NIRI thus 
covered a $22 \times 22$ arcsec field.

	$H$ and $K'$ images were recorded of a region that included the $R=9.5$ 
magnitude star USNO 1575-03026118 (RA = 09:55:35.1, DEC = $+$69:38:56), which 
served as the AO reference beacon for these observations. USNO 1575-03026118 is 
located about 1 arcmin off of the plane of the M82 disk. 
The location of USNO 1575-03026118 with respect to 
the main body of the galaxy is shown in Figure 1.

	Ten 120 sec exposures were recorded through each filter, with two 
exposures recorded at each point in a five-point x-shaped dither pattern. 
The seeing was good (i.e. better than median) and stable when the $K'$ data 
were recorded, and all of the exposures in that filter were used to 
create a final image. However, while the seeing was good when 
the $H$ observing sequence was initiated, conditions quickly deteriorated, 
and only the first 3 $H$ exposures were used in this study.

	The data were processed using a standard pipeline for near-infrared 
images. Each image was divided by a flat-field frame, and the DC sky level 
was subtracted from the result. Interference fringes and the thermal signatures 
of warm elements along the optical path were then removed by subtracting 
a calibration frame that was constructed by combining flat-fielded and DC 
sky-subtracted images of blank sky fields, which were recorded immediately 
following the M82 observations. The construction of this 
calibration frame, and the subsequent removal of 
fringes and thermal artifacts from the science 
data, is not affected by rapid time-scale variations 
in the sky level, as these variations were removed earlier in the processing 
sequence when the DC sky levels were subtracted from the individual science 
and background field images. The images thus processed were 
aligned and median-combined to suppress bad pixels and cosmic rays, and 
then trimmed to the region of full exposure time. The portion of the final $K'$ 
image that is used in the photometric analysis (\S 3.1) is 
shown in Figure 2. The stars in this image have FWHM = 0.08 arcsec, and 
USNO 1575-03026118 is the bright source to the left of the field center.

\section{RESULTS}

\subsection{Photometric Measurements}

	The brightnesses of stars were measured with the point-spread function 
(PSF) fitting routine ALLSTAR (Stetson \& Harris 1988). PSFs, stellar 
co-ordinates, and preliminary brightnesses were obtained 
using tasks in the DAOPHOT (Stetson 1987) photometry package. A 
preliminary pass with the FIND routine identified almost 750 sources in $K'$. 
However, almost 80\% of these were within $\sim 3$ arcsec of 
USNO 1575-03026118, where the PSF wings of this star clearly elevate the noise 
level. A subsequent visual inspection of the detected sources revealed that the 
vast majority of those within 3 arcsec of the guide star were noise features, 
although a handful of real sources were also present. For comparison, at radii 
in excess of 3 arcsec from USNO 1575-03026118 the 
vast majority of detected objects appeared to be real stars. Rather 
than completely abandon the area within 3 arcsec of the guide 
star, it was decided to retain sources detected by eye in both the $H$ 
and $K'$ images in this region. This `by eye' selection procedure introduces an 
obvious bias towards brighter objects, and complicates efforts to estimate 
sample completeness. Consequently, the sources within 3 arcsec of the guide 
star are not included in the discussion of the LF in \S 3.3.

	Anisoplanicity causes the PSF to vary with location across the frame, 
and places a maximum angular distance from the guide star over which 
photometry can be performed. For this study, 
the photometric analysis was restricted to a 
region where it was judged that the PSF was stable, 
based on a visual assessment of PSF shape (McClure et al. 1991) and sharpness. 
Such an analysis indicated that photometry could be performed on stars out to 
8 arcsec from USNO 1575-03026118, and this region is shown in Figure 2. 

	Sample completeness and the random and systematic 
errors in the photometry were estimated by running artificial star experiments, 
in which scaled versions of the PSF, including photon noise, were added to the 
final images in the interval between 3 and 8 arcsec from USNO 1575-03026118. 
These experiments use the single PSF computed for each filter, and so do not 
account for anisoplanicity. Stars were added with $K$ between 20 and 24, and 
$H-K = 0$. The addition of artificial stars to the data elevates the amount of 
crowding. In an effort to keep this effect small, a total of 18 simulations 
were run, with only 20 artificial stars (roughly 10\% of the total number 
of stars actually detected in $K'$ between 3 and 8 arcsec from the guide 
star) added per simulation.

	The results of the artificial star experiments in K' are 
summarized in Figure 3, and it is evident that systematic 
errors in the photometry and sample incompleteness only become significant when 
$K > 23$. The photometric errors predicted by the artificial star experiments 
match the observed scatter in the data (\S 3.2), lending confidence to the 
results from these experiments. The tendency for 
$\Delta$, the mean difference between the 
actual and measured brightness, to increase towards fainter magnitudes occurs 
because faint stars are more likely to be detected if they are situated on 
positive noise spikes, an effect that biases the measured 
brightnesses to higher values. 

\subsection{The $(K, H-K)$ CMD}

	The $(K, H-K)$ CMD of M82 is shown in Figure 4, and 
there is a clear sequence extending from $K = 23$ to $K = 20.0$. 
There are two indications that anisoplanicity does not contribute 
significantly to the scatter in these data. First, the sources within 3 arcsec 
of USNO 1575-03026118, which are plotted as open squares in Figure 4, fall 
along the same general trend as stars at larger radii, which are plotted 
as crosses. Second, the error bar in Figure 4 shows the scatter predicted 
from the artificial star experiments, which used a 
single fixed PSF, and the observed dispersion in $H-K$ 
near $K = 22$ is consistent with that predicted by the 
artificial star experiments. In particular, the standard deviation in $H-K$ of 
stars with $K$ between 21.5 and 22.5 is $\pm 0.109$ mag, whereas the artificial 
star experiments predict a scatter of $\pm 0.084$ mag; an F-test 
confirms that these two values are not significantly different. These 
comparisons indicate that random errors, rather than systematic errors due to 
anisoplanicity, are the dominant source of scatter in the photometry.

	The number of stars in the CMD drops noticeably when 
$K \leq 21.7$, and the presence of a discontinuity at $K = 21.7$ is confirmed 
by convolving the LF, which is discussed in \S 3.3, with a Sobel edge-detection 
kernel. With a distance modulus of 27.95 (Sakai \& Madore 1999), then 
$K = 21.7$ corresponds to M$_K = -6.2$, which is within the range 
of RGB-tip brightnesses expected for old, moderately metal-poor systems (e.g. 
Ferraro et al. 2000). The stars with $K > 21.7$ also have 
$H-K \sim 0$, which is similar to the color of 
RGB stars (e.g. Davidge 2001). Given the similarities 
with the photometric properties of RGB stars, coupled with 
the clear jump in number counts at $K = 21.7$ in the $K$ LF (\S 3.3), we 
conclude that the stars with $K > 21.7$ and $H-K \sim 0$ in Figure 4 
are evolving on the RGB.

	While no members of the Local Group are presently experiencing the same 
level of star formation as M82, comparisons with nearby well-studied systems 
may still provide insight into the stellar content of the M82 ALTAIR field. For 
the present study, comparisons are made with the central region of the dwarf 
elliptical galaxy NGC 205, which was studied with the CFHTIR imager by Davidge 
(2003b), and the outer disk of the Local Group dwarf irregular galaxy NGC 
6822, which was studied by Davidge (2003c) with the CFHT AO system. Both of 
these galaxies have experienced recent episodes of elevated star formation. NGC 
6822 is of particular interest because the rate of star formation in this 
galaxy appears to have increased markedly during recent epochs (Tolstoy et 
al. 2001), which is reminiscent of what is seen in some parts of M82 
(Parmetier et al. 2003; de Grijs, O'Connell, \& Gallagher 2001).
In fact, the RGB content in the outer disk of NGC 6822 is dominated by 
intermediate age objects (Davidge 2003c), while there are also blue main 
sequence stars that are indicative of more recent star 
formation (e.g. de Blok \& Walter 2003; Wyder 2001).

	The $(M_K, H-K)$ CMD of stars near the center of NGC 205 studied by 
Davidge (2003b), and the three NGC 6822 outer disk fields studied by Davidge 
(2003c), are compared with the M82 data in Figure 5. For consistency, M$_K$ was 
computed for each galaxy using distance moduli obtained from the $I-$band 
brightness of the RGB-tip (Lee, Freedman, \& Madore 1993; Gallart, 
Aparicio, \& Vilchez 1996, Sakai \& Madore 1999).

	The CMDs of the M82 ALTAIR field and the two Local Group galaxies are 
similar. The ridgeline of oxygen-rich AGB stars is well-defined on the NGC 205 
and NGC 6822 CMDs, and the majority of stars in the M82 CMD follow a similar 
sequence. Given that the bright red stellar contents of NGC 205 and the outer 
disk of NGC 6822 are dominated by AGB stars, this comparison suggests that the 
brightest red stars in the M82 field are also evolving on the AGB. This 
conclusion is reinforced in \S 3.3, where it is demonstrated that the relative 
number of stars near the bright end of the M82 field with respect to stars on 
the RGB is consistent with the former evolving on the AGB. 

	The peak AGB brightnesses in NGC 205 is higher than in 
M82, but this might simply be a consequence of the 
much larger number of stars in the NGC 205 dataset. In fact, 
after accounting for sample completeness, the three NGC 6822 fields 
combined have only $2\times$ the number of stars with M$_K$ between --7.0 and 
--4 as in the M82 field, and the peak AGB brightness in the NGC 6822 and M82 
CMDs agree to within a few tenths of a mag in M$_K$. 

	C stars dominate the AGB sequences in the LMC and NGC 205 when $H-K > 
0.5$ and M$_K \leq -7$ (e.g. Davidge 2003b, Hughes \& Wood 1990), 
and there is one object in the M82 ALTAIR field that 
has a color and brightness consistent with it being a C 
star. This star is one of the brightest in the field, and is located outside 
of the area that is affected by the PSF wings of USNO 1575-03026118; 
consequently, the photometry should be reliable. The nature of this object 
is discussed further in \S 4.

\subsection{The $K$ LF}

	The LF of sources detected only in $K'$ and located between 3 and 8 
arcsec from the guide star is shown in Figure 6. The LF follows a power-law at 
the faint end, with a discontinuity near $K = 21.7$ due to the onset of the 
RGB (\S 3.2). The LF appears to be flat at the bright end, although 
small number statistics introduce significant scatter above the RGB-tip.

	The $K$ LFs of RGB stars in metal-poor globular clusters 
follow a power-law with an exponent $x = 0.3$ (Davidge 2001). Old stellar 
systems spanning a broad range of metallicities should follow a similar 
relation, as the rate of evolution on the RGB is insensitive to age and 
metallicity (e.g. VandenBerg 1992). Therefore, a power-law with $x = 0.3$ 
was fit to the LF entries with $K$ between 22 and 23.5, and the result 
is shown as a solid line in the lower panel of Figure 6. The LF entries with 
$K < 21.7$ consistently fall below this fitted relation, as 
expected if the RGB-tip occurs near $K = 21.7$.

	Models predict that evolution on the AGB proceeds at a pace that is 
roughly 4 - 5 times faster than on the RGB, and so at a given brightness the 
ratio of AGB to RGB stars should be $\sim 0.2$. The dashed line in 
Figure 6 shows the $x = 0.3$ power law that was fit to the faint end of the LF, 
but with the y-intercept shifted down by log(0.2). The dashed line passes 
through the points at the bright end of the LF, indicating that the relative 
number of sources with $K < 21.7$, measured with respect to those with $K > 
21.7$, is consistent with the brighter stars evolving on the AGB. Such agreement
is consistent with the RGB and AGB stars having evolved from stars that 
were created during the same star-forming episode.

	The composite M$_K$ LF of the three NGC 6822 fields and the M82 ALTAIR 
field are compared in Figure 7. The NGC 6822 data have been scaled to match the 
total number of stars in the M82 LF with M$_K$ between --4 and --7. 
The LFs of the two systems are in general agreement 
when M$_K > -6$. However, there is a tendency for the NGC 6822 LF to 
consistently fall above that of M82 when M$_K < -6$. This is due at least in 
part to differences in the $K-$band brightnesses of the 
RGB-tip, M$_{K}^{RGBT}$, as M$_{K}^{RGBT} = -6.5$ in NGC 6822 (Davidge 2003c), 
while M$_{K}^{RGBT} = -6.2$ in M82 assuming the Sakai \& Madore (1999) distance.

	M$_{K}^{RGBT}$ is sensitive to both metallicity and age, in 
the sense that it becomes fainter towards lower ages (e.g. Girardi et al. 
2002), and lower metallicities. It is unlikely that the difference in 
M$_{K}^{RGBT}$ is due to metallicity, as NGC 6822 is an intrinsically fainter 
system than M82, and hence should have a lower mean metallicity. Rather, 
the difference in M$_{K}^{RGBT}$ may indicate that the RGB stars in the M82 
ALTAIR field are younger than those in NGC 6822. Davidge (2003c) concluded 
that the RGB stars in the outer disk of NGC 6822 have 
an age of 3 Gyr. The models discussed by Girardi et al. (2000) predict 
that a 0.3 mag drop in M$_{K}^{RGBT}$ at a fixed metallicity occurs when 
$\Delta$(t$_{yr}) \sim 0.5$, and so the RGB stars in M82 would have an 
age $\frac{3.0}{3} \sim 1.0$ Gyr if they have the same metallicity as the RGB 
stars in the outer disk of NGC 6822.

	We caution that the age infered from the RGB-tip is 
uncertain, as the metallicity of the RGB stars in this region of M82 is not 
known, and there are uncertainties in the model physics. Nevertheless, 
the estimated age is consistent with a peak in the star formation rate in 
other parts of the disk that has been associated with the last major 
interaction with M81 (Parmentier et al. 2003; de Grijs et al. 2003). Thus, 
this comparison suggests that the outer disk of M82 may have actively formed 
stars at roughly the same time as other parts of the galaxy, 
and this is consistent with the brightnesses of bright 
AGB stars in this field (\S 4). Given the evidence 
for very recent star formation in the disk on the same side of the center of 
M82 as the ALTAIR field (e.g. Gallagher \& Smith 1999; Smith \& Gallagher 
2003), deeper imaging of this field at visible wavelengths may reveal evidence 
of even younger stars. The detection of a young population would not be 
unexpected, as atomic and molecular gas have been detected outside of the disk 
plane (e.g. Taylor, Walter, \& Yun 2001), and so material to fuel star 
formation is available. 

\section{DISCUSSION AND SUMMARY}

	Blending is a potential source of concern with these data, as fainter 
stars in crowded environments can combine to form an object that will appear 
as a brighter single star. In particular, unresolved pairs of objects on 
the top 0.75 mag of the RGB will blend to form objects that are brighter than 
the RGB-tip. Sakai \& Madore (1999) argued that blending of this nature 
may contribute significantly to the population of stars above the RGB-tip 
in their F555W and F814W WFPC2 observations of a disk edge field on the 
other side of M82 from the ALTAIR field, with the result that they were  
not able to conclude if a bright AGB sequence was present. 

	The upper portions of the AGB and RGB on near-infrared CMDs are more 
extended along the magnitude axis than on CMDs at visible and 
red wavelengths, and this reduces the incidence of blending. 
In fact, there are two indications that blending is not an issue 
with these data. First, the relative numbers of AGB and RGB stars match those 
predicted by models (\S 3.3), suggesting that the brightest 
stars in the ALTAIR field are not blended RGB stars. Second, 
the number density of bright RGB stars is relatively low. There are 23 stars 
in our data within the top 0.75 mag in $K$ of the RGB, and these are located 
over 200 arcsec$^{2}$. If each resolution element has a radius equal to 
one half of the FWHM of the PSF (i.e. 0.04 arcsec), then the density 
of upper RGB stars is only $6 \times 10^{-4}$ per resolution 
element. The incidence of blending is thus negligible, and 
we conclude that the M82 ALTAIR field contains AGB stars that are 
1.7 mag in $K$ brighter than the RGB-tip. In fact, given the 
modest field sampled by ALTAIR, it is likely that the AGB-tip 
is brighter than the brightest AGB star in this field, so that M$_K^{AGB-tip} 
< -8$.

	The presence of AGB stars above the RGB-tip alone is not 
ironclad evidence of an intermediate age population, as the brightness of 
the AGB-tip at a fixed age depends on metallicity, and even moderately 
metal-poor old systems can have a bright AGB component. For example, the 
brightest AGB stars in the [Fe/H] = --0.7 globular cluster 47 Tuc, which has an 
old age (e.g. Hesser et al. 1987), have M$_K \leq -7.2$ when viewed near the 
peak of their light curves, making them roughly 1 mag in $K$ brighter than the 
RGB-tip (Frogel, Persson, \& Cohen 1981). 

	To estimate the age of the AGB stars in the ALTAIR field it is thus 
necessary to know the metallicity of this field. The most 
direct way to estimate the metallicity is to study 
the photometric properties of RGB stars, and this will require high angular 
resolution multi-color observations spanning a broader range of wavelengths 
than the current $H$ and $K'$ data. Insight into the metallicity of the 
ALTAIR field can also be gleaned from studies of other parts of the galaxy. 
The youngest populations in M82 appear to have 
roughly solar metallicities. Martin (1997) 
concluded that [O/H] may be as high as solar in the interstellar medium 
of M82 based on the strengths of various emission lines, while Gallagher 
\& Smith (1999) were able to model the integrated spectra of two young star 
clusters by assuming a solar metallicity. However, the metallicity of 
M82 has evolved with time. Parmentier et al. (2003) computed the metallicities, 
ages, and extinctions of clusters in M82 Region B, and they find (1) a 
broad range of metallicities at all ages, and (2) 
evidence that this area experienced a major enrichment episode following the 
most recent interaction with M81, which occured 1 Gyr in the past.

	The cluster metallicities computed by Parmentier et al. (2003) can be 
used to gain insight into the chemical enrichment history of M82 Region B. In 
Table 1 we show the mean metallicities of clusters in various age intervals, 
computed from the data in Tables 1 and 2 of Parmentier et al. (2003). The 
uncertainties are the errors in the mean. The means 
computed for clusters in regions B1 and B2 separately are not 
statistically different from each other, and so only the mean metallicities for 
clusters in both regions are shown in Table 1. The entries in Table 1 show 
that (1) there is a trend for $\overline{log(z/z_{\odot})}$ to increase 
towards younger ages, and (2) the clusters tend to be moderately metal-poor 
when log(t) $> 9.0$.

	M82 Region 2 samples the disk plane of the galaxy, and studies of 
edge-on late-type galaxies indicate that metallicity drops with vertical 
distance from the disk plane (e.g. Dalcanton \& Bernstein 2002). Therefore, 
the Region B metallicities are upper limits to those expected in the M82 ALTAIR 
field. Based on the data in Table 1, if the M82 ALTAIR field is dominated by 
very old stars then there is a reasonable expectation that it will have a 
metallicity below that of 47 Tuc; however, there are 3 stars in the 
ALTAIR field with M$_K \leq -7.2$, and this is not consistent with an old, 
moderately metal-poor population. On the other hand, if the ALTAIR field 
is dominated by intermediate age stars then it will have a higher metallicity, 
and a brighter AGB-tip than is seen in 47 Tuc, and this interpretation is 
consistent with the observed AGB content. Therefore, based on the brightnesses
of AGB stars and the brightness of the RGB-tip (\S 3.3), we conclude that 
the ALTAIR field is not dominated by an old population.

	We close by noting that an object with photometric properties that 
are consistent with it being a C star has been detected in the M82 ALTAIR 
field. Clearly, caution should be exercised when dealing with only one object, 
as there is the real possibility that this object may be a foreground or 
background source, and the reality of a C star population in the outer disk of 
M82 can only be confirmed with a survey covering a much larger area.
The detection of C stars in the outer disk of M82 would be 
of interest, as it would allow for additional insight into the 
age of this part of the galaxy. In particular, C star formation appears to 
cease at an age of $4 - 5$ Gyr in systems with metallicities comparable to 
the LMC, and at progressively younger ages as metallicity 
increases (e.g. Cole \& Weinberg 2002). Models predict 
that the contribution made by C stars to the total AGB light depends on 
age, peaking at $t = 1$ Gyr in moderately metal-poor systems, 
where they contribute 14\% of the light; this contribution drops to 7\% 
when $t = 6$ Gyr, and to zero when $t > 6$ Gyr (Maraston 1998). 
These numbers are consistent with the C star content of NGC 205 (Davidge 
2003b) and intermediate age clusters in the LMC (Maraston 1998).
Therefore, if the outer disk of M82 has an age of a few Gyr, and is moderately 
metal-poor, then a large C star population would be expected. 
This being said, it is interesting that the candidate C 
star detected in the ALTAIR field accounts for 10\% of 
the total AGB $K-$band brightness, which is consistent 
with what is expected from a moderately metal-poor intermediate age population. 

\clearpage
\parindent=0.0cm

\begin{table*}
\begin{center}
\begin{tabular}{lc}
\tableline\tableline
log(t$_{yr}$) & $\overline{[Z/Z_{\odot}]}$ \\
\tableline
8.0 -- 8.5 & $-0.1 \pm 0.1$ \\
8.5 -- 9.0 & $-0.3 \pm 0.2$ \\
9.0 -- 9.5 & $-0.7 \pm 0.2$ \\
9.5 -- 10.0 & $-1.2 \pm 0.2$ \\
\tableline
\end{tabular}
\end{center}
\caption{Mean Metallicities of Clusters in M82 Region B}
\end{table*}

\clearpage

\begin{center}
FIGURE CAPTIONS
\end{center}

\figcaption[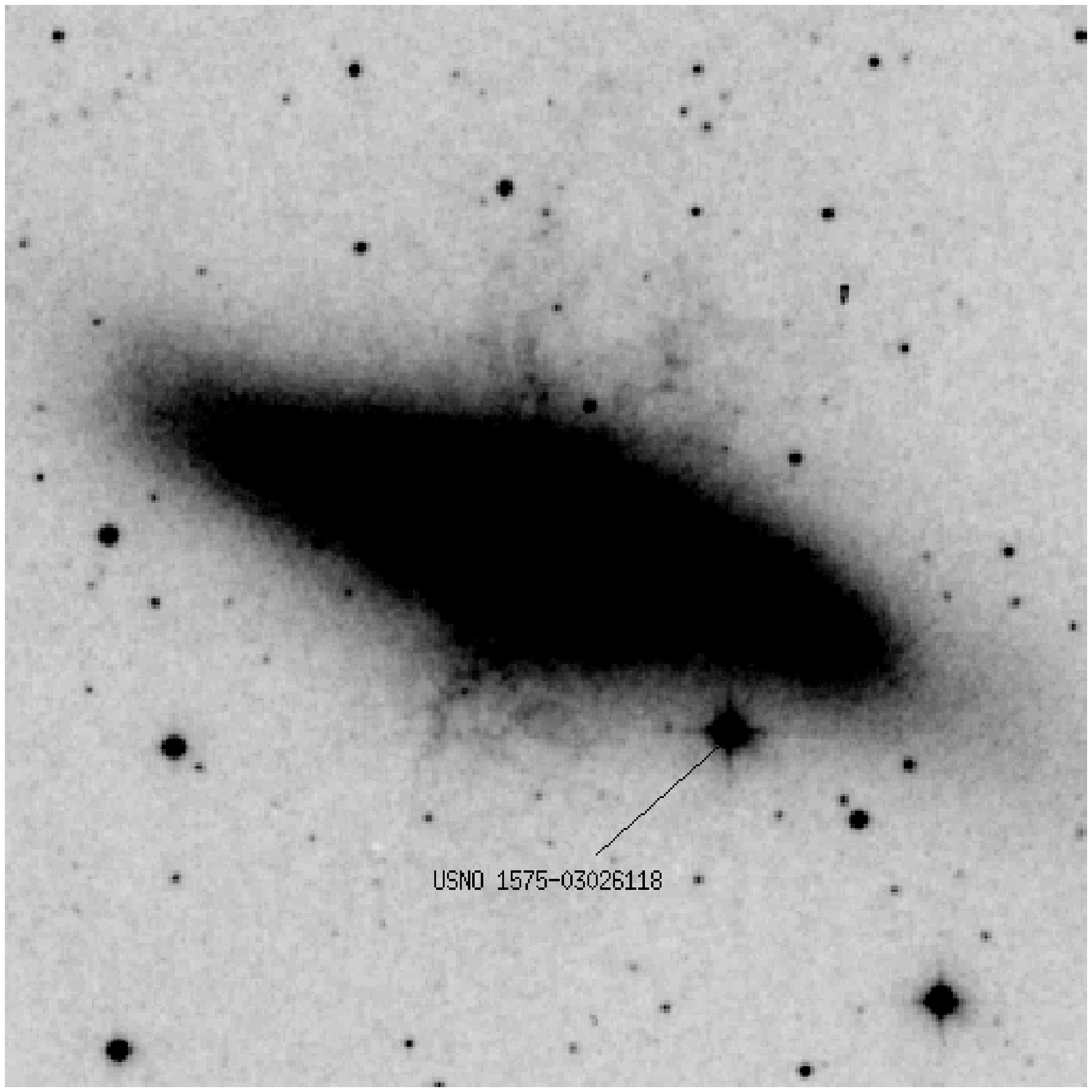]
{A $10 \times 10$ arcmin portion of the DSS showing the location of USNO 
1575-03026118, which was the AO guide star for these observations. The field 
imaged by ALTAIR $+$ NIRI is comparable in size to the stellar disk of USNO 
1575-03026118 in this figure. North is at the top, and East is to the left.}

\figcaption[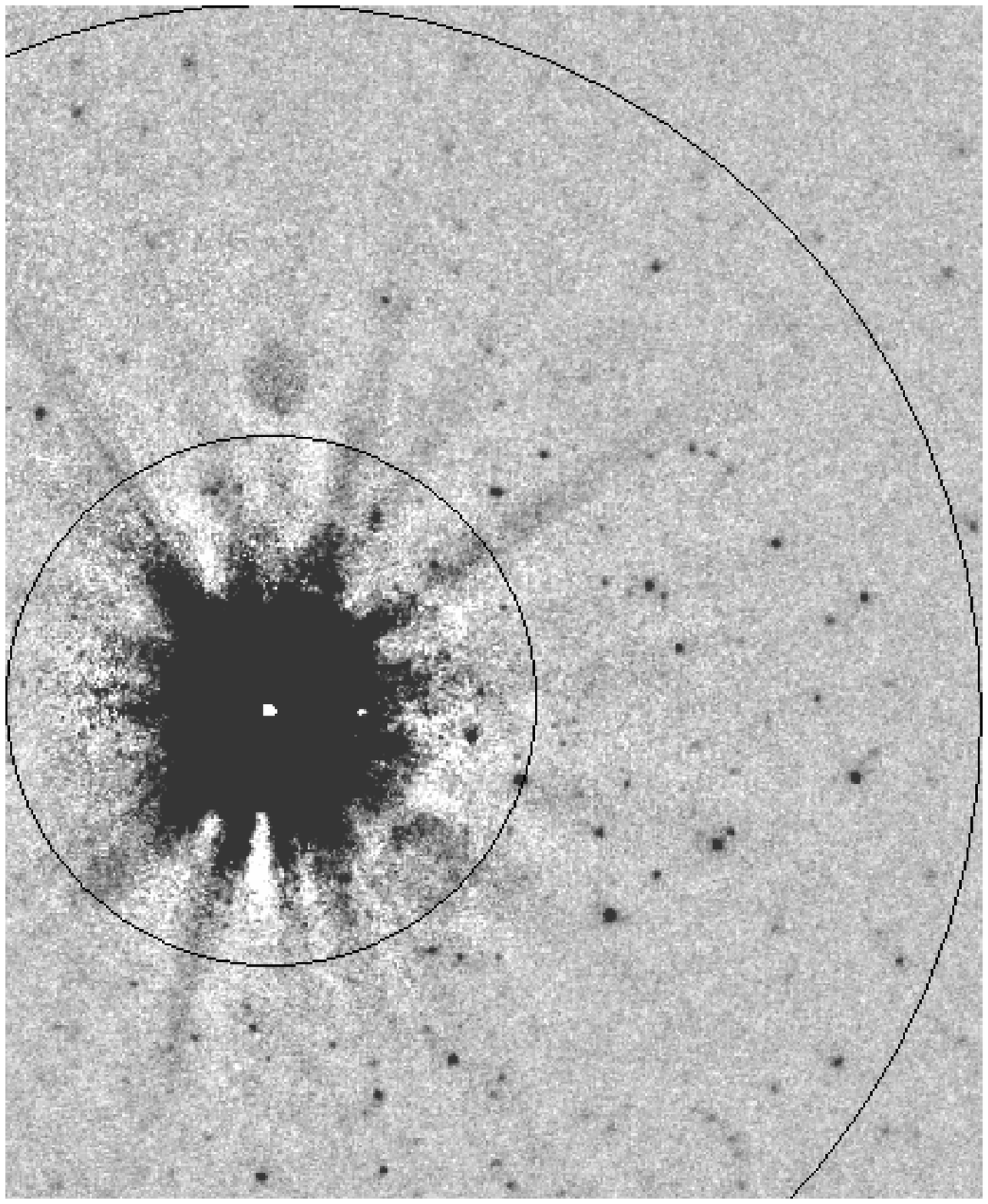]
{The portion of the final $K'$ image of the M82 field that was used in the 
photometric analysis. The inner circle has a radius of 3 arcsec, and only 
sources that were clearly detected by eye in both filters and were not affected 
by diffraction features were photometered within this circle. The outer circle, 
which has a radius of 8 arcsec, defines the maximum distance out to which 
the PSF was judged to be stable. The radial coverage is not complete because 
USNO 1575-03026118 was not centered on the detector.}

\figcaption[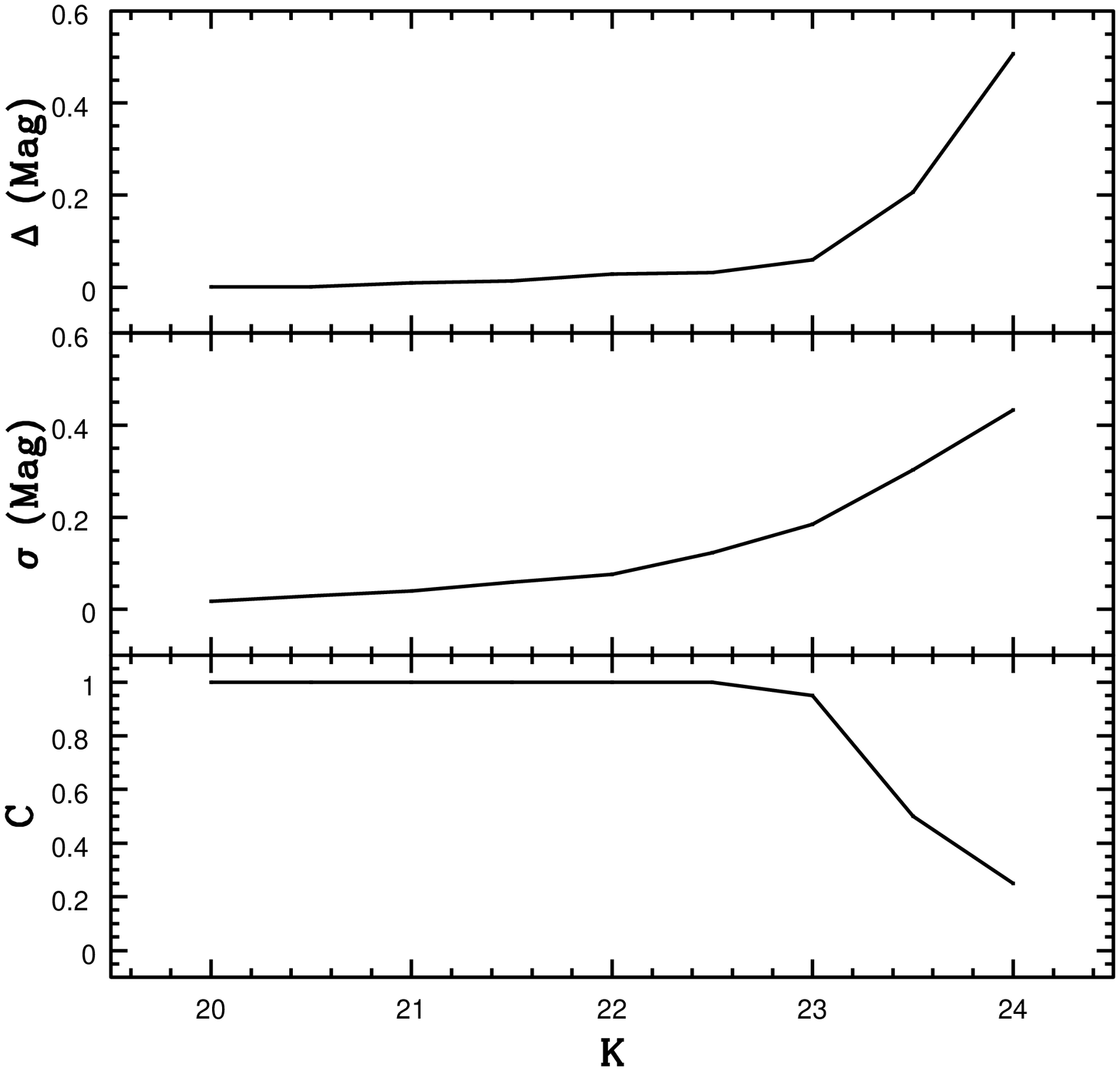]
{The results of artificial star experiments conducted with the K' image. 
Artificial stars were inserted into the processed $K'$ image at radial 
distances between 3 and 8 arcsec from USNO 1575-03026118. $\Delta$ is the mean 
difference between the actual and measured brightnesses of the artificial 
stars, while $\sigma$ is the standard deviation of 
the values used to compute $\Delta$ at each magnitude. $C$ is the 
completeness fraction, which is the ratio of the number of artificial stars 
detected to the number actually inserted. Note that sample incompleteness and 
systematic errors in the photometry only become significant when $K > 23$.} 

\figcaption[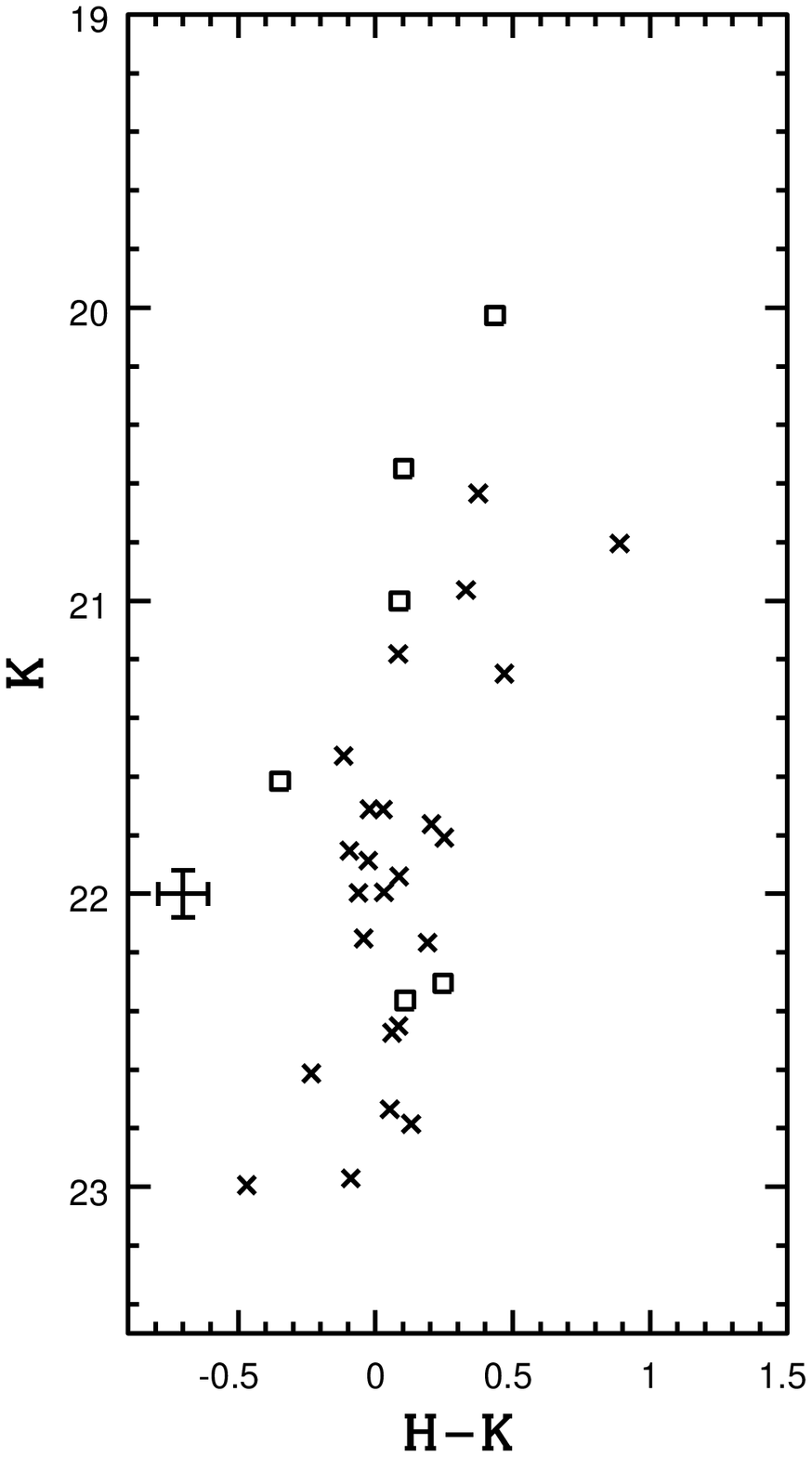]
{The $(K, H-K)$ CMD of the M82 ALTAIR field. 
Stars that are more than 3 arcsec from USNO 1575-03026118 are 
plotted as crosses, while those within 3 arcsec of this star 
are plotted as open squares. The error bars show the uncertainty in the CMD at 
$K = 22$ due to random errors, as predicted by the artificial star experiments. 
The observed dispersion along the color axis is 
comparable to that predicted by the artificial star experiments.} 

\figcaption[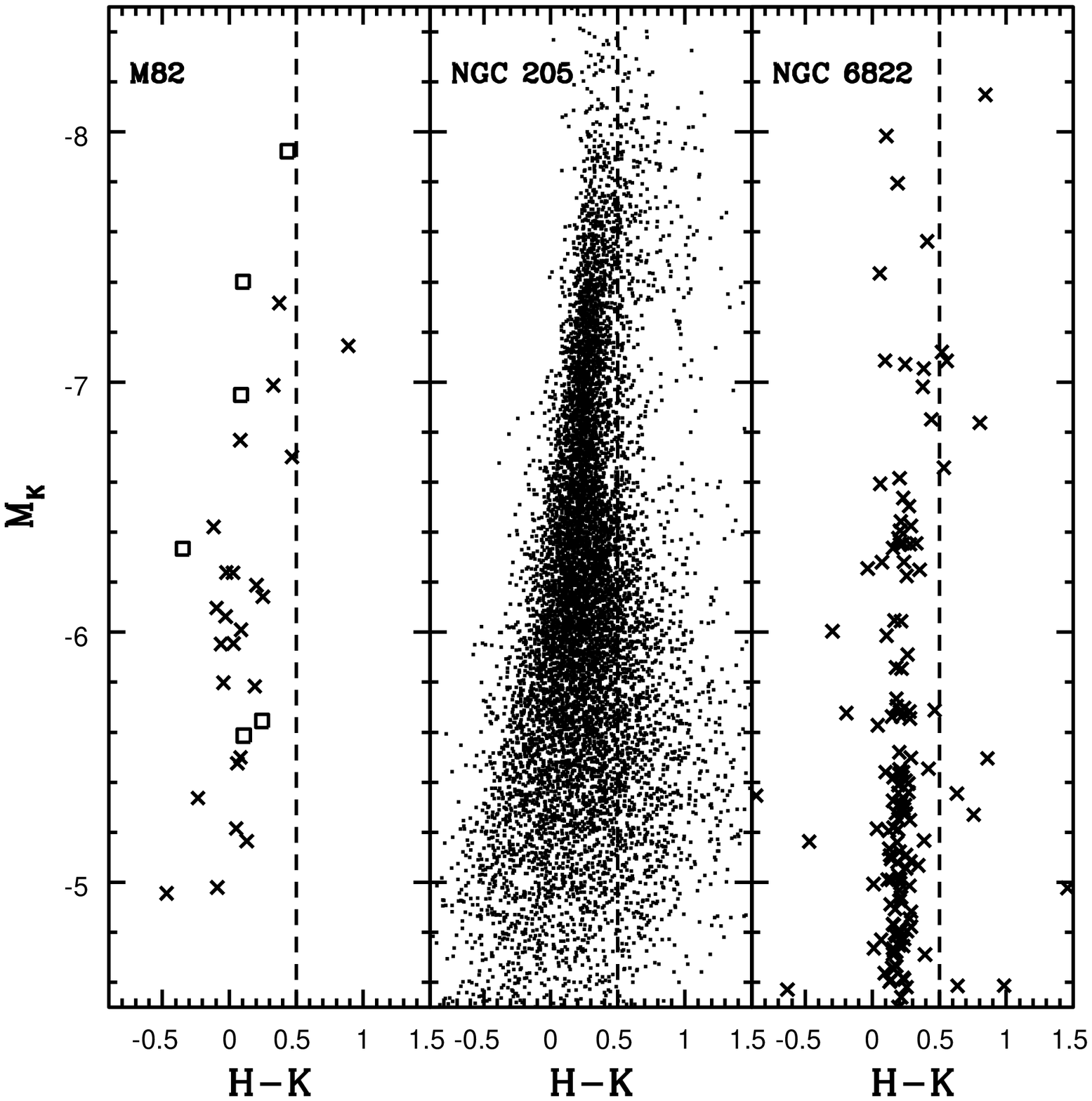]
{The $(M_K, H-K)$ CMD of the M82 ALTAIR field is shown in the left 
hand panel. Stars at distances in excess of 3 arscec from USNO 1575-03026118 
are plotted as crosses, while those located within 3 arcsec of this star 
are plotted as open squares. The middle panel shows the composite 
$(M_K, H-K)$ CMD of the NGC 205 inner and outer regions studied by 
Davidge (2003b), while the right hand panel shows the composite 
$(M_K, H-K)$ CMD of the three NGC 6822 outer disk fields studied by Davidge 
(2003c). For consistency, M$_K$ was computed for all three galaxies using the 
$I-$band RGB-tip brightness. The dashed lines mark $H - K = 0.5$, which is 
the color threshold for the onset of C stars in the LMC (Hughes \& Wood 1990) 
and NGC 205 (Davidge 2003b).} 

\figcaption[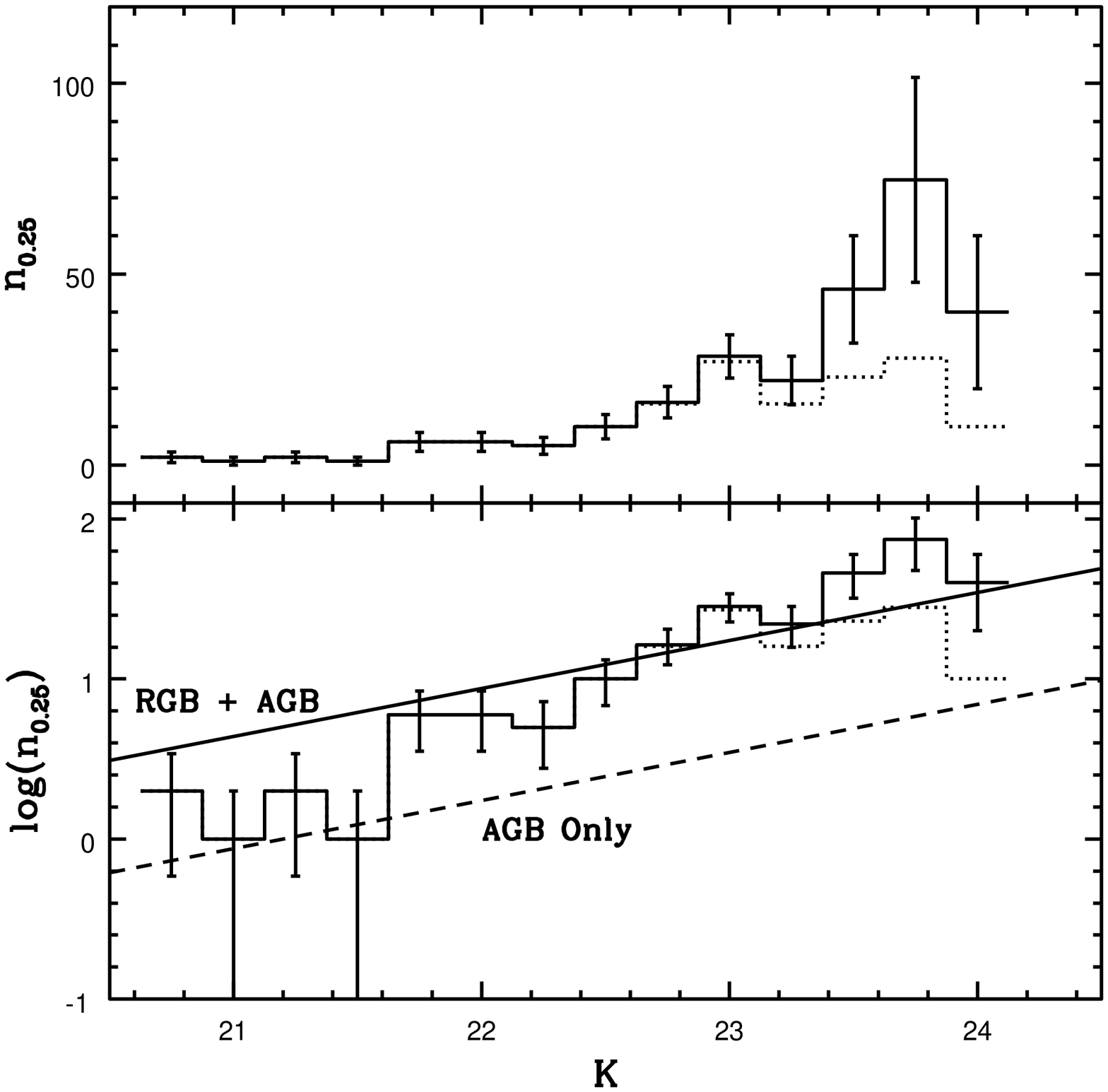]
{The LF of stars between 3 and 8 arcsec of USNO 1575-03026118 
that were detected only in $K'$. n$_{0.25}$ is the number of stars 
per 0.25 mag interval in $K$. The raw LF is plotted as a dotted line, 
while the completeness-corrected LF is plotted as a solid line. The 
error bars include counting statistics and the uncertainties in the completeness
corrections. The line labelled `RGB $+$ AGB' in the lower panel shows a least 
squares fit of a power-law with a fixed exponent $x = 0.3$ to the entries with 
$K$ between 22 and 23.5. The dashed line in the lower panel marked `AGB Only' 
shows the same relation but with the y-intercept shifted down by log(0.2). Note 
that the latter line passes through the data points at the bright end, 
indicating that the ratio of bright ABG stars to RGB stars is consistent with 
predictions from stellar evolution models.} 

\figcaption[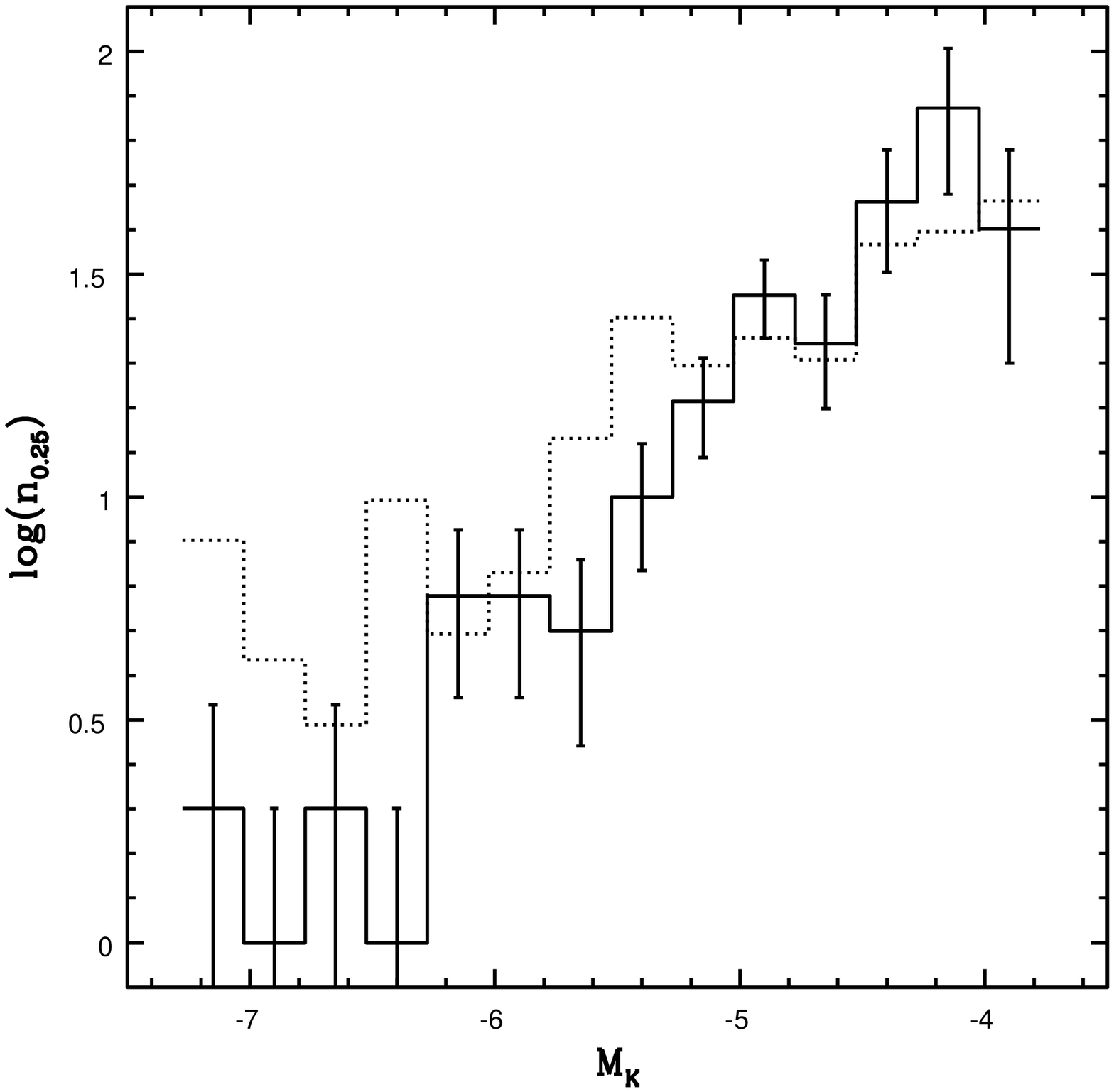]
{The completeness-corrected M$_K$ LF of the M82 ALTAIR field 
compared with the composite LF of the three outer disk fields in NGC 
6822 discussed by Davidge (2003c). The M82 LF is the solid line, 
while the NGC 6822 LF is the dotted line. n$_{0.25}$ is the number of 
stars per 0.25 mag interval in M$_K$, and the error bars show the 
uncertainties due to photon statistics and the completeness corrections.
The NGC 6822 LF has been shifted along the vertical axis to 
match the number of points in M82 with M$_K$ between --7 and --4. Note the 
excess number of stars with M$_K < -6.2$ in NGC 6822 when compared with M82.} 
\end{document}